 \documentstyle[preprint,epsf]{jpsj}

\def\runtitle{A quantum Monte Carlo algorithm realizing an intrinsic relaxation}
\def\runauthor{Tota {\sc Nakamura} and Yoshiyuki {\sc Ito}}

\title{A quantum Monte Carlo algorithm
realizing an intrinsic relaxation
}

\author{Tota {\sc Nakamura} and Yoshiyuki {\sc Ito}}

\inst{Department of Applied Physics, Tohoku University,
         Aoba, Sendai, Miyagi  980-8579}

\recdate{~~~~~~~~~~~~~~~~~~~~~~~~~~~~~~}

\abst{
We propose a new quantum Monte Carlo algorithm
which realizes a relaxation intrinsic to the original quantum system.
The Monte Carlo dynamics satisfies the dynamic scaling relation 
$\tau\sim \xi^z$
and is independent of the Trotter number.
Finiteness of the Trotter number just appears as the finite-size effect.
An infinite Trotter number version of the algorithm is also formulated,
which enables us to observe
a true relaxation of the original system.
The strategy of the algorithm is a compromise between 
the conventional worldline local flip
and the modern cluster loop flip.
It is a local flip in the real-space direction and
is a cluster flip in the Trotter direction.
The new algorithm is tested by the transverse-field Ising model in two dimensions.
An accurate phase diagram is obtained.
}

\kword{quantum Monte Carlo method, nonequilibrium relaxation method,
quantum dynamics, transverse-field Ising model}

\begin{document}
\sloppy
\maketitle

The Monte Carlo method is a powerful tool to investigate the 
condensed matter physics.
The method directly benefits from a progress of microprocessors.
It is also suitable for the parallel computing.
Developments of various Monte Carlo algorithms consist of an important part
of the statistical physics in these decades.
\cite{binderMC}

The quantum version is called
a quantum Monte Carlo (QMC) method.\cite{qmcbook}
A $d$-dimensional quantum system is decomposed 
into a $(d+1)$-dimensional classical system,
where the actual simulations are performed.
The additional dimension is called the Trotter direction and its length
is denoted by the Trotter number $m$.
The original quantum system is recovered 
in the limit of the infinite Trotter number.

The first update algorithm in QMC is the worldline local flip.
A worldline, which is an outcome of the local spin conservation,
is moved locally in the $(d+1)$-dimensional lattice. 
An acceptance ratio of this flip becomes worse 
and the Monte Carlo dynamics freezes
as the Trotter number increases to approach the original quantum system.
This is purely a technical problem not from a physical origin as the
critical slowing down.
Recently, the loop flip algorithm in QMC 
\cite{loopalgo1,loopalgo2,loopalgo3,continuousIT} 
is developed in order to solve this problem.  
It is also possible to take 
the infinite Trotter number limit beforehand. \cite{continuousIT}
A nonlocal loop is defined in the $(d+1)$-dimensional lattice for the update.
All spins on the loop are simultaneously flipped.
This loop corresponds to the correlated spins in this $(d+1)$-dimensional
lattice.
Therefore, the algorithm drastically reduces the correlation time.

The loop algorithms, or generally cluster algorithms, 
are a trend of recent developments in Monte Carlo methods.
It accelerates the Monte Carlo simulations suffering from the slow dynamics.
Another trend is the nonequilibrium relaxation (NER)
method.\cite{ner1,ner2,ner3}
The NER method positively makes use of the critical slowing down
to detect the phase transition.
One observes a relaxation function of a physical quantity at each temperature.
If it exhibits an algebraically slow relaxation, 
the temperature is judged to be a critical point.
The method works very well in the slow-dynamic systems.
\cite{totaner1,totaner2,totaner3}
A Monte Carlo investigation on the relaxation process is
a good approach to the critical phenomena.

An updating algorithm in the NER simulation must realize an
{\it intrinsic relaxation} of the system.
For example, an algebraic relaxation should be observed at the critical point.
A relaxation in the paramagnetic phase should be exponential with a
finite correlation time corresponding to a finite correlation length.
The requirement is guaranteed by updating degrees of freedom with
the same length at each time step.
One realization is a single spin flip algorithm in the classical 
spin systems.

In regard to the QMC method, however,
an algorithm which realizes a relaxation intrinsic to the original
quantum system has not been invented yet.
A relaxation function by the loop algorithm has nothing to do with the physics.
The NER analysis is possible by using the worldline local flip.\cite{nonomura}
A ratio between the inverse temperature $J/T$ 
and the Trotter number $m$ is fixed finite.
The simulation exhibits a relaxation 
of the $(d+1)$-dimensional classical system.
It suffers from the freezing as the Trotter number increases.

Aim of this Letter is to propose a new QMC algorithm realizing the 
relaxation intrinsic to the original quantum system.
A shortcoming of the worldline local flip algorithm is
a mixture of a physical relaxation and an unphysical relaxation
causing the freezing.
The new algorithm extracts only a physical relaxation part.
The unphysical one is eliminated by using an idea of the loop algorithm.
The infinite Trotter number version is also formulated.

A basic idea of the present algorithm is 
to make an update {\it local in the real-space direction} and
{\it global in the Trotter direction}.
The former one is to ensure the Monte Carlo dynamics reflecting the
physics of the original quantum system.
We typically choose one real-space interaction bond or 
one real-space spin as a local unit of the updating.
One Monte Carlo step corresponds to a time unit in which
a change of this real-space length occurs.
The latter idea is realized by a cluster flip 
extending only in the Trotter direction.
Each updating unit is connected along the Trotter direction
in a same manner as the Swendsen-Wang algorithm \cite{swendsenwang} 
in one dimension.
A cluster of the connected updating units is flipped simultaneously.
A length of the cluster corresponds to the correlation in the Trotter direction.
Therefore, the freezing due to the Suzuki-Trotter
decomposition is solely eliminated.
The present update algorithm can be considered as 
a {\it single quantum-spin flip}.

\begin{figure}[htb]
\begin{center}
 \epsfxsize = 8.5cm
 \epsffile{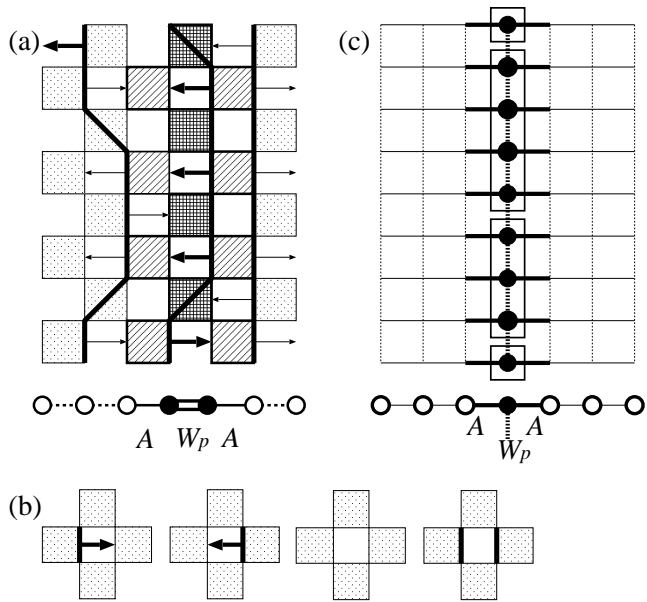}
\end {center}
 \caption {
(a)The Suzuki-Trotter decomposition of the Heisenberg chain.
The worldlines are depicted by bold lines.
Pseudo-spins are depicted by arrows.
Those updated in this trial are depicted by bold arrows.
(b) Definitions of a pseudo-spin in a blank plaquette due to the worldline
configuration.
No pseudo-spin is assigned to 
the worldline vacancy and the double occupancy.
(c) The Suzuki-Trotter decomposition of the transverse-field Ising chain.
Spins updated in this trial is depicted by solid circles.
Rectangles indicate clusters.
 \label{fig:lattice}
}
\end{figure}

We consider an $S=1/2$ Heisenberg chain for a simple
explanation of the algorithm.
\begin{equation}
{\cal H}= \sum _{i}
J \mbox{\boldmath $S$}_{i} \cdot \mbox{\boldmath $S$}_{i+1}
\end  {equation}

The system is decomposed into a checkerboard plane as shown in 
Fig.~\ref{fig:lattice}(a).
An updating unit is four spins consisting of
a blank plaquette in this figure.
We define a pseudo-spin in a blank plaquette as 
depicted in Fig.~\ref{fig:lattice}(b).
If a worldline runs vertically in the left (right) side of the plaquette,
a pseudo-spin pointing to the right (left) is assigned.
If there is no vertical worldline or there are two worldlines in both sides,
no pseudo-spin is assigned.
A pseudo-spin represents a possible worldline move.
It changes the direction, if the spin flip is accepted.
The worldline local flip is regarded as a single spin flip of this pseudo-spin.

An actual procedure of the updating is as follows.
We choose one real-space bond to try an update
(the double-line bond connecting two solid spins 
in Fig.~\ref{fig:lattice}(a)).
Pseudo-spins in regard with these two real-space spins are updated.
(Bold arrows in the figure.)
A Boltzmann weight $W$ participating in this update is 
written as a product of two parts.
\begin{equation}
W = W_p \times A.
\end  {equation}
One consists of weights of plaquettes on the updating bond 
(\epsfysize=0.50cm \epsffile{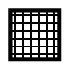})
as denoted by $W_p$.
These plaquettes can be considered as effective bonds
connecting pseudo-spins on the updating bond.
They are used to define a cluster of pseudo-spins.
The other part consists of weights of plaquettes on the neighboring bonds 
(\epsfysize=0.50cm {\epsffile{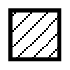}})
as denoted by $A$.
The weights determine a probability to accept the cluster flip.

If two pseudo-spins neighboring along the Trotter direction point to the same
direction,
they are connected to form a cluster
with a probability $1-p$ of 
\begin{equation}
p\equiv \epsfysize=0.5cm {\epsffile{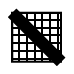}}
       /\epsfysize=0.5cm {\epsffile{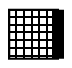}}
      =\tanh (J/mT ) \sim J/mT.
\end{equation}
Pseudo-spins in a cluster are flipped simultaneously with a probability
calculated from the weights of plaquettes on the neighboring bonds $A$.
A product of the weights of plaquettes adjacent to each
pseudo-spin in the cluster is calculated before ($A_{\rm initial}$)
and after ($A_{\rm final}$) the flip.
We accept the flip with a probability
\begin{equation}
\frac{A_{\rm final}}{A_{\rm initial}+A_{\rm final}}.
\label{eq:A}
\end{equation}
We try this flip for each cluster independently.
A new worldline configuration is obtained by new pseudo-spin
configurations.

The ergodicity of this update algorithm 
is same as the worldline local flip.
The worldline global flip is necessary to change the magnetization.
The present algorithm is an adoption of the Swendsen-Wang algorithm
under the external field.
Here, a molecular field from the neighboring real-space spins
is regarded as the external field.
Therefore, 
the detailed balance is also satisfied as is guaranteed in the
Swendsen-Wang algorithm.
An important notice is that the acceptance of a flip 
does not depend on the Trotter number nor the temperature.
An average size of the cluster is about $1/p \sim mT/J $.
Then, the contribution from the neighboring interaction bonds $A$ is
an order of 
$
\exp\left[ \frac{J}{mT}\times \frac{1}{p}\right] 
\sim O(1).
$
The freezing due to the decomposition is solely eliminated by this relation.
A cluster analysis is easy because it is performed only in one dimension.
It enables us to write a fast program 
without difficulties for various systems.
An actual computational time of one Monte Carlo step is 
almost same as that of the conventional local flip one.

It is possible to take the infinite Trotter number limit beforehand.
In this scheme
we introduce a `breakup', which is a domain wall of the pseudo-spins,
i.e., where a worldline hops.\cite{loopalgo1,loopalgo2,loopalgo3}
A location of a breakup is stored in memory to define a state.
A probability of a breakup to exist is $p\sim J/mT$.
An average number of the breakups is $J/T$ in the infinite $m$ limit.
Therefore, 
among the pseudo-spins on a selected updating bond we put
breakups by Poisson random numbers with the expectation number $J/T$.
The pseudo-spins between the neighboring breakups are 
flipped by using the probability of Eq.~(\ref{eq:A}).
Here, we do not discriminate between the newly-assigned breakups 
and the already-existing breakups.

An updating procedure for the transverse-field Ising model is almost
same as for the Heisenberg model.
Only a difference is that there is no local spin conservation in this model.
An updating unit, which is a plaquette in the Heisenberg model,
shrinks to a single spin. 
Therefore, the single-spin flip is possible in the $(d+1)$-dimensional lattice.
The Hamiltonian is
\begin{equation}
{\cal H}=\sum_{\langle i,j \rangle} JS_i^z S_j^z -\sum_i \Gamma S_i^x.
\end  {equation}
The Suzuki-Trotter decomposition is shown in Fig. \ref{fig:lattice}(c).
A Boltzmann weight is assigned to each interaction bond.
First, we choose one real-space spin to try an update.
Spins along a line in the Trotter direction will be updated.
(Solid circles in Fig. \ref{fig:lattice} (c).)
A cluster is defined by connecting two spins neighboring 
in the Trotter direction with a probability $1-p$ of 
\begin{equation}
p=\tanh(\Gamma/mT)\sim \Gamma/mT,
\end{equation}
if they take same spin value.
The clusters are shown by rectangles in Fig. \ref{fig:lattice}(c).
They are flipped with a probability of Eq. (\ref{eq:A}).
The weights $A_{\rm initial}$ and $A_{\rm final}$ are calculated by
molecular fields 
from the spins neighboring in the real-space direction
to the cluster.
The infinite Trotter number version in this model is also trivial.
We put breakups with the expectation number $\Gamma/T$ along a line in
the Trotter direction of the updated spin.
A cluster between two breakups is flipped by Eq. (\ref{eq:A}).

As a demonstration of the new algorithm we consider the transverse-field
Ising model in two dimensions.
The classical phase transition occurs at a temperature which 
varies with $\Gamma/J$.
The quantum phase transition occurs at $T=0$ for a finite $\Gamma/J$ value.

\begin{figure}[htb]
\begin{center}
 \epsfxsize = 8.5cm
 \epsffile{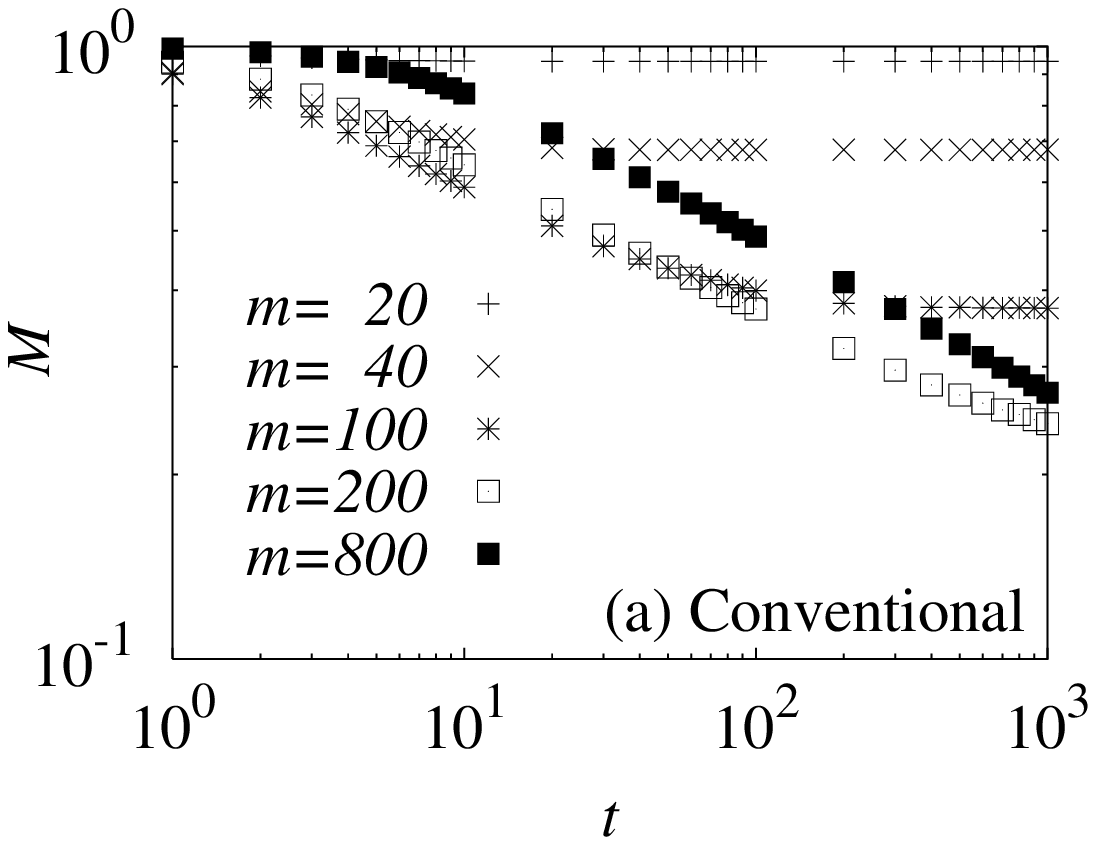}

 \epsfxsize = 8.5cm
 \epsffile{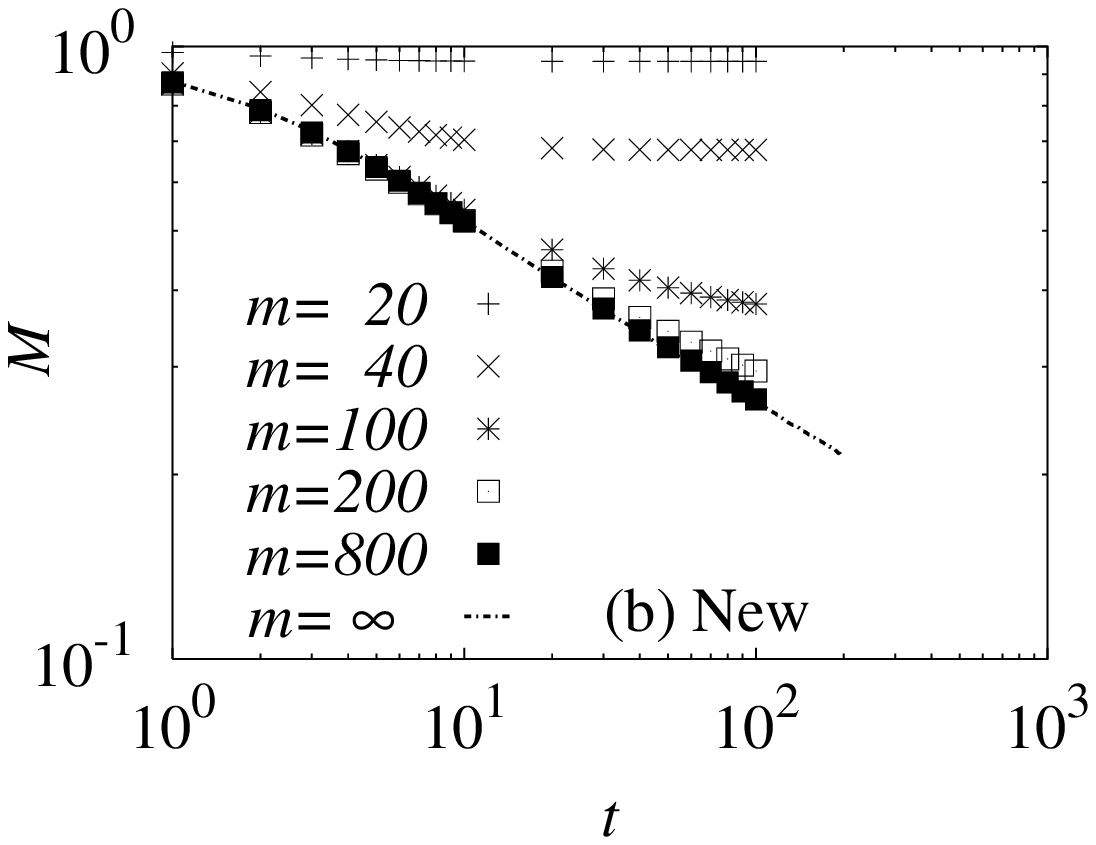}
\end {center}
 \caption {
  NER plots of the magnetization of 2D
transverse-field Ising model near the critical point. 
$\Gamma/J=3.05$, $T/J=0.1$. The real-space system size is $400\times 400$.
(a) By the conventional local flip algorithm.
(b) By the new flip algorithm proposed in this Letter.
Data of the infinite Trotter number version ($m=\infty$)
is depicted by a line.
 \label{fig:nama}
          }
\end  {figure}

Figure \ref{fig:nama} shows NER plots of the magnetization $M$
at $T/J=0.1$ and $\Gamma/J=3.05$.
This point is in the vicinity of the transition point.
The relaxation functions depend on the Trotter number
in the conventional algorithm as shown in Fig.~\ref{fig:nama}(a).
As the Trotter number increases, 
it becomes later for the algebraic relaxation to begin.
On the other hand, the relaxation functions by the new algorithm 
essentially do not depend on the Trotter numbers as shown 
in Fig.~\ref{fig:nama}(b).
The relaxation function of the infinite Trotter number is depicted by a line.
The figure clearly shows that the
Trotter number dependence only appears as the finite-size effect.
Before it appears, the relaxation coincides with that of the infinite 
Trotter number.
The relaxation of the original quantum system is considered to be realized.
The equilibrium data of small Trotter numbers
are consistent between two algorithms.
The logarithmic slope of the relaxation $-\beta/z\nu$
also coincides with each other.
Therefore, we can conclude that two algorithms are observing
the same critical phenomenon.
A computational time for one Monte Carlo step is almost same between
two algorithms.
The total necessary computation time to achieve the same resolution
is reduced to $1/10$.

\begin{figure}[htb]
\begin{center}
 \epsfxsize = 8.5cm
 \epsffile{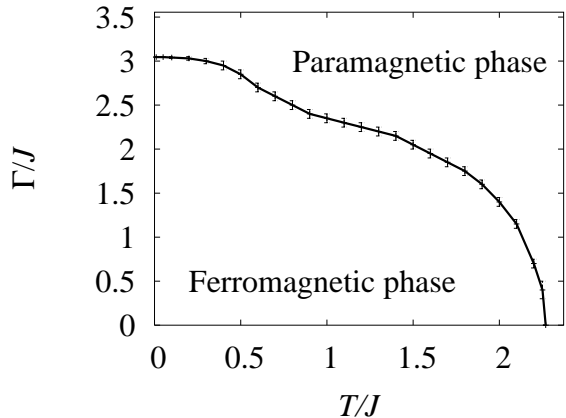}
\end {center}
 \caption {
A phase diagram of 2D transverse-field Ising model obtained by
the NER using the infinite Trotter number algorithm.
The real-space system size is $199\times 200$ ($T/J \ge 0.1$) and
$149\times 150$ ($T/J=0.01$).
The time range is 1000 steps.
\label{fig:phasediagram}
}
\end  {figure}

Figure \ref{fig:phasediagram} shows a phase diagram obtained by the 
NER analysis of the magnetization using the infinite Trotter number
version of the present algorithm.
The smallest value of $\Gamma$ at which the relaxation exhibits 
the exponential decay is the lower bound of the paramagnetic phase.
The largest value of $\Gamma$ at which the relaxation converges 
to a finite value is the upper bound of the ferromagnetic phase.
The phase boundary line is in between.
The lowest temperature we have carried out the simulation is $T/J=0.01$.
Within the system size ($149\times 150$) and the time range (1000 steps)
of the simulation at this temperature there is no temperature effect and the 
data can be considered as those of the ground state.
Our estimate of the quantum transition point is $\Gamma_{\rm c}=3.044(2)$.
In the classical limit ($\Gamma/J=0$) the phase transition of the 2D classical
Ising model occurs at $T/J=2.269$.
We have also verified this temperature and the critical exponents by using
the same program.
The phase diagram is obtained very accurately compared with previous
investigations.\cite{ikegami}
The shape of the phase diagram
qualitatively agrees with the experimental result
of LiHoF$_4$, \cite{bitko}
which is considered to realize the transverse-field Ising model
in three dimensions.

\begin{figure}[htb]
\begin{center}
 \epsfxsize = 8.5cm
 \epsffile{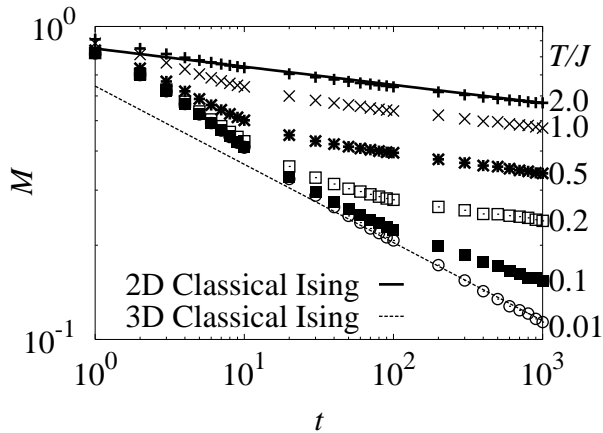}
\end {center}
 \caption {
  An NER plot of the magnetization on the phase boundary line.
  The temperature is as denoted outside the plot.
The corresponding $\Gamma$ values are 
$\Gamma/J=1.39$($T/J=2.0$),
$2.34$($1.0$),
$2.845$($0.5$),
$3.02$($0.2$),
$3.038$($0.1$), and
$3.044$($0.01$).
  Lines depict $t^{-\beta/z\nu}$ with exponents of 2D classical
Ising model and 3D classical Ising model.
The real-space system size is $199\times 200$ for $T/J=2.0-0.1$ and
$149\times 150$ for $T/J=0.01$.
 \label{fig:exponent}
          }
\end  {figure}

Figure \ref{fig:exponent} shows an NER plot on the phase boundary line
when changing the temperature.
The logarithmic slopes $-\beta/z\nu$ converge to the 
same value in the long time limit at high temperatures.
It indicates that they belong to the same universality class.
The slope is consistent with a choice of $\beta=1/8$, $\nu=1$, and $z=2.16$,
which is the exponents of the 2D classical Ising model.
As the temperature decreases, 
it takes a longer time to observe the 2D classical Ising behavior.
In the low temperature limit ($T/J=0.01$) the relaxation exhibits 
the universality of the 3D classical Ising model, which is 
equivalent to the quantum phase transition of the present model.
The slope is consistent with the exponents
$\beta=0.325$, $\nu=0.635$, and $z=2.05$.\cite{ito3Dising}
We observe in this figure a systematic crossover from the 2D Ising to
the 3D Ising universality.
A relaxation function traces the trail of the ground-state 
($T/J=0.01$) until the temperature effect appears.
Then, 
it changes the behavior to the 2D classical Ising universality.

In summary, 
a quantum Monte Carlo algorithm realizing the  relaxation 
intrinsic to the original quantum system is formulated.
Analyses on the relaxation of a quantum system are done without an 
ambiguity of the Suzuki-Trotter decomposition.
This is made possible by updating single-quantum-spin degrees of freedom
as the updating unit.
Since a length scale updated in one Monte Carlo step is definite,
the algebraic relaxation is expected to be realized in the critical phase.
It is an open question whether or not
the dynamics of this algorithm explains a time evolution of 
a quantum system which strongly couples with the heat reservoir.
It is expected that the use of the present algorithm helps a progress of
the field.
We remark here that the negative-sign problem is not solved by this algorithm.
Therefore,
applications to the various transverse-field Ising models 
and to the non-frustrated Heisenberg/XY models are fruitful.

The author TN thanks a financial support from The
Sumitomo Foundation.
The use of random number generator RNDTIK programmed by
Professor N.~Ito and Professor Y.~Kanada is gratefully acknowledged.

\end{document}